\documentclass[aps,prb,twocolumn,groupedaddress]{revtex4-1}

\usepackage{graphicx}
\usepackage{color}

\begin{document}


\title{Magnon dispersion shift in the induced-ferromagnetic phase of the noncentrosymmetric MnSi}


\author{Taku J Sato}
\email[]{taku@tagen.tohoku.ac.jp}
\affiliation{Institute of Multidisciplinary Research for Advanced Materials, Tohoku University, 2-1-1 Katahira, Sendai 980-8577, Japan}

\author{Daisuke Okuyama}
\affiliation{Institute of Multidisciplinary Research for Advanced Materials, Tohoku University, 2-1-1 Katahira, Sendai 980-8577, Japan}

\author{Tao Hong}
\affiliation{Quantum Condensed Matter Division, Oak Ridge National Laboratory, Oak Ridge, TN 37831, USA}

\author{Akiko Kikkawa}
\affiliation{RIKEN Center for Emergent Matter Science (CEMS), Wako, Saitama 351-0198, Japan}

\author{Yasujiro Taguchi}
\affiliation{RIKEN Center for Emergent Matter Science (CEMS), Wako, Saitama 351-0198, Japan}

\author{Taka-hisa Arima}
\affiliation{Department of Advanced Materials Science, University of Tokyo, Kashiwa, Chiba 277-8561, Japan}
\affiliation{RIKEN Center for Emergent Matter Science (CEMS), Wako, Saitama 351-0198, Japan}

\author{Yoshinori Tokura}
\affiliation{RIKEN Center for Emergent Matter Science (CEMS), Wako, Saitama 351-0198, Japan}
\affiliation{Department of Applied Physics and Quantum Phase Electronics Center (QPEC), University of Tokyo, Tokyo 113-8656, Japan}

\date{\today}

\begin{abstract}
Small angle neutron inelastic scattering measurement has been performed to study the magnon dispersion relation in the field-induced-ferromagnetic phase of the noncentrosymmetric binary compound MnSi.
For the magnons propagating parallel or anti-parallel to the external magnetic field, we experimentally confirmed that the dispersion relation is asymmetrically shifted along the magnetic field direction.
This magnon dispersion shift is attributed to the relativistic Dzyaloshinskii-Moriya interaction, which is finite in noncentrosymmetric magnets, such as MnSi.  
The shift direction is found to be switchable by reversing the external magnetic field direction.
\end{abstract}


\maketitle

\section{Introduction}

Symmetry is a ubiquitous concept that determines degeneracy of energy states of particles in crystalline materials.
Exemplified by electrons, degeneracy for both $\uparrow$- and $\downarrow$-spin states and $+\vec{q}$ and $-\vec{q}$ states is expected in a centrosymmetric crystal under time-reversal symmetry, {\it i.e.}, $E(q, \uparrow) = E(-q, \downarrow)$ and $E(q, \uparrow) = E(q, \downarrow)$.
While time-reversal-symmetry breaking results in the trivial ferromagnetic band splitting, spatial-inversion-symmetry breaking leads to spin splitting in a non-trivial manner due to the relativistic spin-orbit coupling; $E(q, \uparrow) \neq E(-q, \uparrow)$ but still $E(q, \uparrow) = E(-q, \downarrow)$.
A well-known example of such asymmetry-induced band splitting is the Rashba effect~\cite{RashbaEI60,BychkovYA84}, which has been experimentally confirmed in various noncentrosymmetric systems~\cite{Bihlmayer2015,Manchon2015,LaShellS96,NittaJ97,BellRL62,IshizakaK11}.

The above symmetry constraint should hold not only for real particles but also for quasi-particles in crystalline materials.
A ferromagnetic magnon is one of such quasi-particles, and has a symmetric and quadratic dispersion relation in centrosymmetric compounds, {\it i.e.}, $E(q) = D_{\rm s}q^2$, where $D_{\rm s}$ stands for the spin stiffness.
Upon breaking the inversion symmetry, it supposedly becomes asymmetric, $E(q) \neq E(-q)$.
The simplest asymmetric dispersion may be:
\begin{equation}\label{eq:dispersion}
E(q) = E_0 + D_{\rm s} (q - q_0)^2,
\end{equation}
where the bottom of the original quadratic dispersion relation shifts to a finite value $q_0$.
($E_0$ stands for a gap energy.)
This shifted quadratic magnon dispersion relation has been known theoretically for decades, and was microscopically attributed to the antisymmetric Dzyaloshinskii-Moriya (DM) interaction originating from the relativistic spin-orbit coupling~\cite{MelcherRL73,KataokaM87}.

Although theoretically well understood, experimental confirmation of the magnon dispersion shift has been largely limited to ferromagnetic thin films~\cite{ZakeriKh10,Di2015,Stashkevich2015,Nembach2015}, where the inversion symmetry is trivially lost.
For magnons in bulk noncentrosymmetric ferromagnets, very recently, the nonreciprocal propagation of the magnons was detected in the two compounds LiFe$_5$O$_8$~\cite{IguchiY15} and Cu$_2$OSeO$_3$~\cite{SekiS15} in a GHz (or $\mu$eV) range using a modern microwave resonance technique.
The nonreciprocal propagation of magnons is due to different frequencies for the $+q$ and $-q$ magnons, and hence is a sign of asymmetry in the magnon dispersion relation.
It should be noted, however, that microwave resonance observes magnons in a $q \simeq 0$ region, where the group velocity can be negative, depending on an effective sample thickness, due to dipole-dipole interactions~\cite{StencilDD93}; indeed, the backward-propagating mode, called magnetostatic mode, dominates in a thin sample.
For a thicker sample, another issue, a distribution of resonance frequencies along the sample thickness direction, prohibits unique assignment of the magnon frequency.
Hence, the magnon dispersion in a microscopic length scale, where the exchange and DM interactions certainly dominates, is experimentally inconclusive in bulk noncentrosymmetric ferromagnets.
Neutron inelastic scattering is a common tool to measure magnon dispersion relations in bulk magnets, nonetheless, one could hardly find experimental observation of the asymmetric magnon dispersion.
This is due to stringent requirements for directions of the external magnetic field, magnon propagation vector, and crystallographic axis, and also due to the extremely high momentum-transfer ($Q$) and energy ($E$) resolutions necessary to detect the dispersion shift.

MnSi is a prototypical itinerant chiral helimagnet and has been studied for half a century~\cite{Williams66}.
MnSi belongs to the noncentrosymmetric space group $P2_13$, where the finite DM interaction is expected.
In the zero external magnetic field, a single-$q$ helical magnetic structure is stabilized below the transition temperature $T_{\rm c} \approx 29.5$~K.
The helical structure has a quite long modulation period, characterized by the modulation vector $\vec{q} \approx (0.016, 0.016, 0.016)$ (r.~l.~u.)~\cite{IshikawaY76,IshidaM85}.
By applying the external field, the single-$\vec{q}$ helical structure first transforms into the conical structure, and then into the induced-ferromagnetic structure~\cite{IshikawaY84}.
This compound attracts renewed interests because of the recent observation of the skyrmion-lattice phase under finite external magnetic field in vicinity of $T_{\rm c}$~\cite{MuhlbauerS11}.
Low-energy spin excitations in MnSi have been also studied in detail.
In the zero-field helical phase, infinitely folded helimagnon bands were observed, and were attributed to its incommensurability~\cite{JanoschekM10}.
In the induced-ferromagnetic phase, well-defined magnon excitations were observed at low energies $E < 2.5$~meV, whereas for $E > 2.5$~meV the magnon excitation becomes overdumped due to the particle-hole excitation continuum, called Storner continuum~\cite{IshikawaY77,BoniP11}.
Related to the magnon dispersion shift, there have been two pioneering works in literature~\cite{Shirane1983,SokoloffJB1984}.
They used the {\it polarized-neutron} inelastic scattering technique, and selectively observed single-handed spiral magnetic correlation.
Either slight asymmetry in the inelastic scattering spectrum~\cite{Shirane1983}, or possible slight $Q$-shift of the magnon peak position by reversing field direction~\cite{SokoloffJB1984}, was reported there.
Nonetheless, because of the insufficient $Q$- and $E$-resolutions, the magnon band shift was only speculative, being far from conclusive.
Quite recently an another trial was made to experimentally study the magnon band shift using polarized small angle neutron scattering technique~\cite{Grigoriev15}.
In the experiment, polarization dependence of the energy integrated diffraction intensity was recorded, and from its $\vec{Q}$-space asymmetry the magnon dispersion shift was inferred.
It should be, however, pointed out that the used technique is in principle energy insensitive, and hence cannot be a direct observation of magnon dispersion relation.
Apparently, an inelastic experiment is necessary.

In this work, we performed small angle neutron inelastic scattering using {\it unpolarized neutrons} to confirm the long-sought magnon dispersion shift in the induced-ferromagnetic phase of the noncentrosymmetric MnSi.
It was found that the magnon peak appears only in either $\hbar\omega > 0$ or $< 0$ side of the inelastic spectrum, depending on the direction of the external magnetic field.
As far as we know, this clear asymmetry in the {\it unpolarized neutron} excitation spectra is the first observation of this kind, nonetheless, is a ubiquitous property of inelastic spectra for noncentrosymmetric magnets.
The magnon dispersion relation was experimentally determined; the dispersion shift was unambiguously detected, with the shift direction and its magnitude being in quantitative agreement with the theoretical prediction.
This confirms the nonreciprocality of the magnon propagation in the noncentrosymmetric ferromagnet in the microscopic length scale.

\section{Experimental}
\begin{figure}
\includegraphics[scale=0.4, angle=-90, trim={4cm 3.1cm 5cm 3cm}]{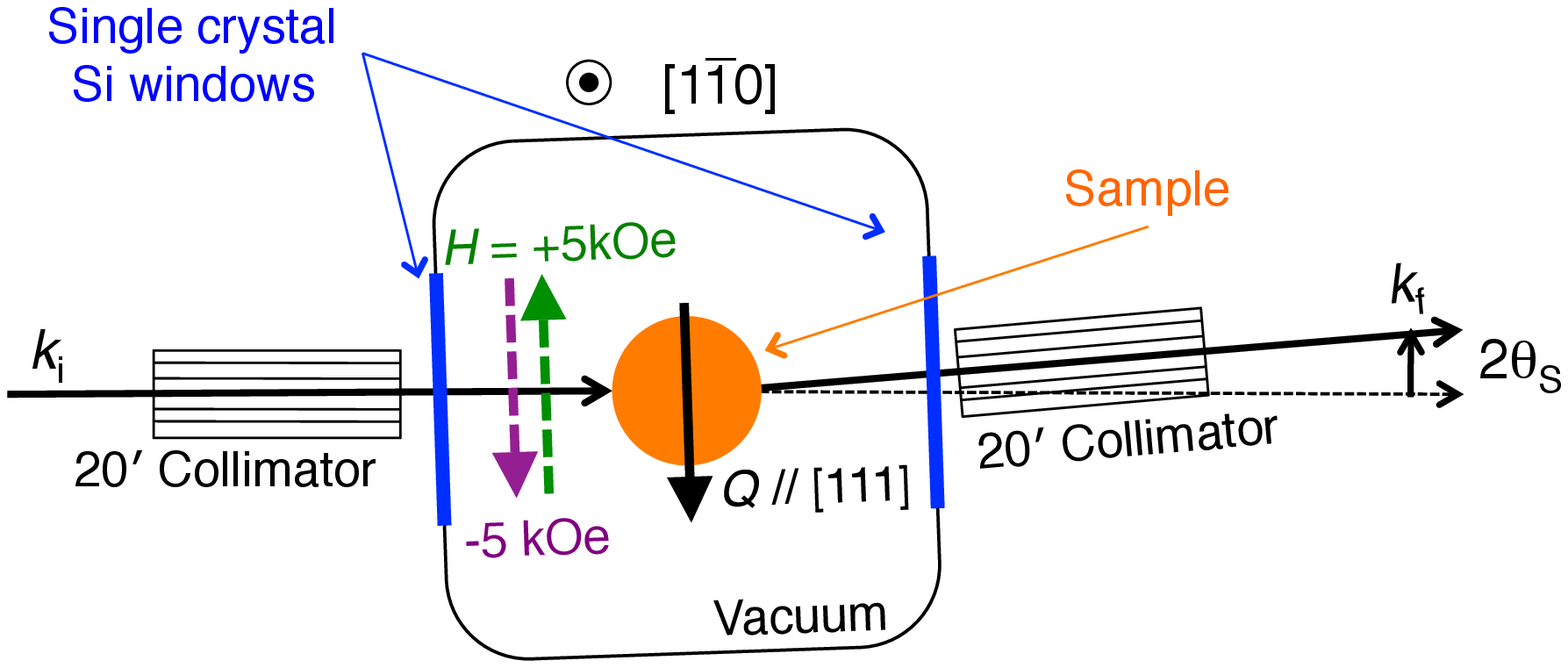}
\caption{(Color online) Schematic illustration of the experimental setup for the small angle neutron inelastic measurements.
The external magnetic field was fixed anti-parallel ($H > 0$) or parallel ($H < 0$) to the momentum transfer $\vec{Q}$, which was set along the crystallographic $[111]$ direction.}\label{fig1}
\end{figure}

Single crystals of MnSi were grown by the Bridgman method using Al$_2$O$_3$ crucibles sealed in quartz tubes~\cite{Stishov08}.
Two crystals are co-aligned with the total mass $\sim$ 18~g.
The neutron inelastic scattering experiment was performed using the cold-neutron triple-axis spectrometer CTAX, installed at the HFIR reactor of Oak Ridge National Laboratory.
Pyrolytic graphite 002 reflections were used for both the monochromator and analyzer.
Cooled Be filter was placed after the sample to eliminate higher harmonic neutrons.
The co-aligned crystals were loaded in the horizontal field superconducting magnet with the $hhl$ plane set to the scattering plane, and with the magnetic field $\vec{H}$ antiparallel to $\vec{Q} \parallel [111]$.
($\vec{Q}$ is defined as $\vec{Q} = \vec{k}_{\rm i} - \vec{k}_{\rm f}$.)
The ferromagnetic magnons were measured around the origin $\vec{Q} = 0$, using the small outgoing neutron energy $E_{\rm f} = 3.25$~meV and tight collimations of $20^{\prime}$ placed before and after the sample.
The resulting $Q$-resolution was $\Delta Q \simeq 0.01$~\AA$^{-1}$, whereas the energy resolution $\Delta E \simeq 0.1$~meV at the elastic position.
For the inelastic scattering, background in the low-$Q$ region is mainly due to the small angle scattering from window materials and air, and was reduced by using a large vacuum chamber around the sample with single crystal Si windows.
Remaining background was estimated by the base temperature scans performed at $T \simeq 4$~K, where the magnon excitations were strongly suppressed due to the small Bose population factor.
For all the inelastic spectra shown in this report, the estimated background was removed from the raw spectra observed at higher temperatures.
The experimental setup is schematically illustrated in Fig.~1. 

\section{Results and Discussion}
\begin{figure}
  \includegraphics[scale=0.35,angle=-90]{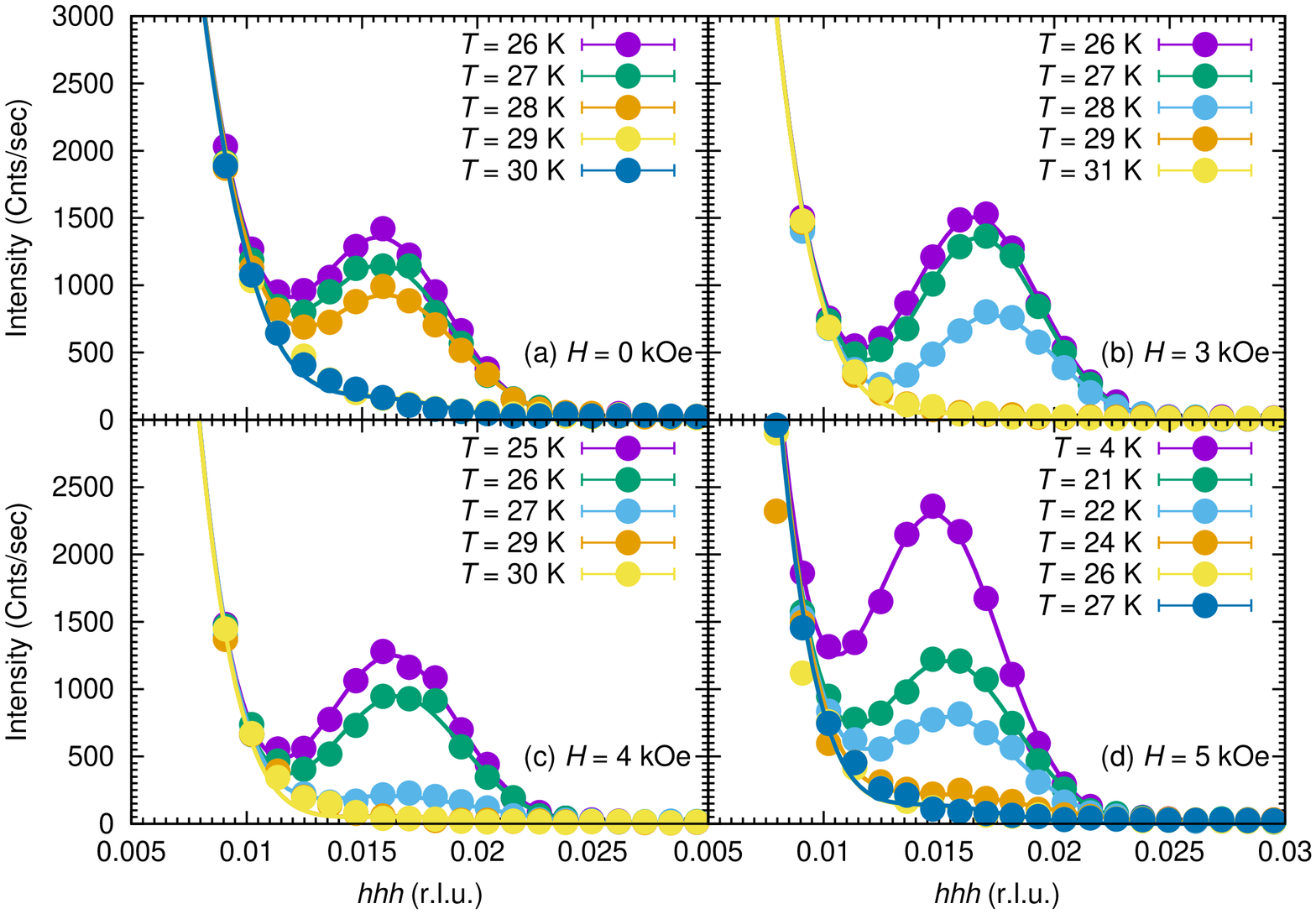}\\
  \includegraphics[scale=0.25,angle=-90]{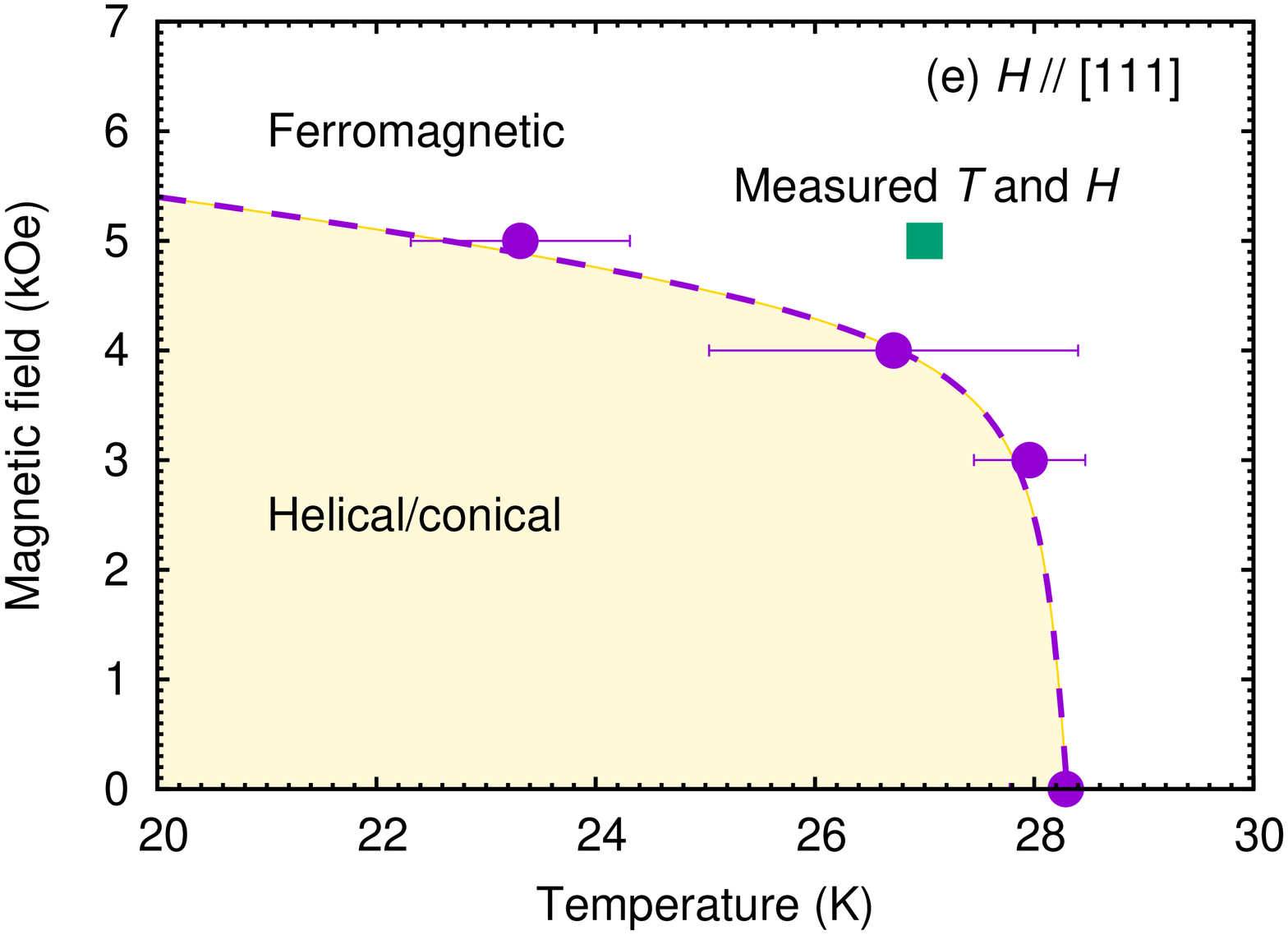}
  \caption{(Color online) (a-d) Neutron diffraction patterns observed under $H = 0, 3, 4$ and 5~kOe.
Solid lines are the fitting results using gaussian functions.
(e) Phase boundary for the helical/conical and induced-ferromagnetic phases in MnSi for $H \parallel [111]$.
Filled circles stand for the estimated transition temperatures.
Filled square indicates the $T$ and $H$ point, where the present small angle inelastic neutron scattering measurements were performed.
}
\end{figure}

First, we checked the phase boundary between the helical/conical phase and paramagnetic/induced-ferromagnetic phase using elastic neutron scattering.
Figure~2(a) shows the temperature dependence of the neutron diffraction pattern under the zero external field.
The incommensurate peak was clearly observed at $\vec{Q} = (0.016, 0.016, 0.016)$ (r.~l.~u.) at low temperatures, while it disappears above $29$~K.
From the temperature dependence of the peak intensity, the transition temperature was estimated as $T_{\rm c} = 28(1)$~K, which is consistent with the earlier report~\cite{MuhlbauerS11}.
The diffraction patterns under the finite magnetic fields $H = 3, 4$, and 5~kOe are shown in Figs.~2(b-d).
The incommensurate peak was observed at low temperatures for all $H$, and $T_{\rm c}$ was similarly estimated from their temperature dependence.
The estimated $T_{\rm c}$ is shown in Fig.~2(e).
$T_{\rm c}$ decreases quickly as $H$ increases, again being in reasonable agreement with the earlier bulk measurement~\cite{KadowakiK82}.
One may note that the peak position shifts to lower $Q$ under the higher magnetic field $H = 5$~kOe at low temperatures.
This is due to the temperature dependence of the modulation vector; indeed the peak position is in good agreement with the earlier report~\cite{Grigoriev06} by taking account of its temperature dependence.

Next, we measured the neutron inelastic spectra at $T = 27$~K under the magnetic field $H = +5$~kOe or $H = -5$~kOe, where the system is certainly in the induced-ferromagnetic phase as evidenced from the above diffraction experiment.
The inelastic spectrum observed at the momentum transfer $\vec{Q} = (0.05, 0.05, 0.05)$~(r.~l.~u.) under $H = 5$~kOe is shown in Fig. 3(c).
Surprisingly, the inelastic peak appears only at $\hbar\omega = -0.18$~meV in the negative energy side, which corresponds to the process where neutrons gain energy from the system.
The inelastic spectrum under the reversed magnetic field $H = -5$~kOe is shown in Fig.~3(d).
Clearly, the magnon peak appears in the positive energy side at $\hbar\omega = 0.18$~meV for this field direction; again asymmetry in the inelastic spectrum is observed.
Figures 3(a), 3(b), 3(e), and 3(f) represent inelastic spectra at different $\vec{Q}$ positions.
At all the $\vec{Q}$ positions the asymmetric appearance of the magnon excitation peak has been confirmed.
This is the first demonstration of asymmetry in the {\it unpolarized-neutron} inelastic-scattering spectrum for noncentrosymmetric ferromagnets, as far as we know.

\begin{figure}
\includegraphics[scale=0.48, angle=-90, trim={0cm 3cm 0cm 0cm}]{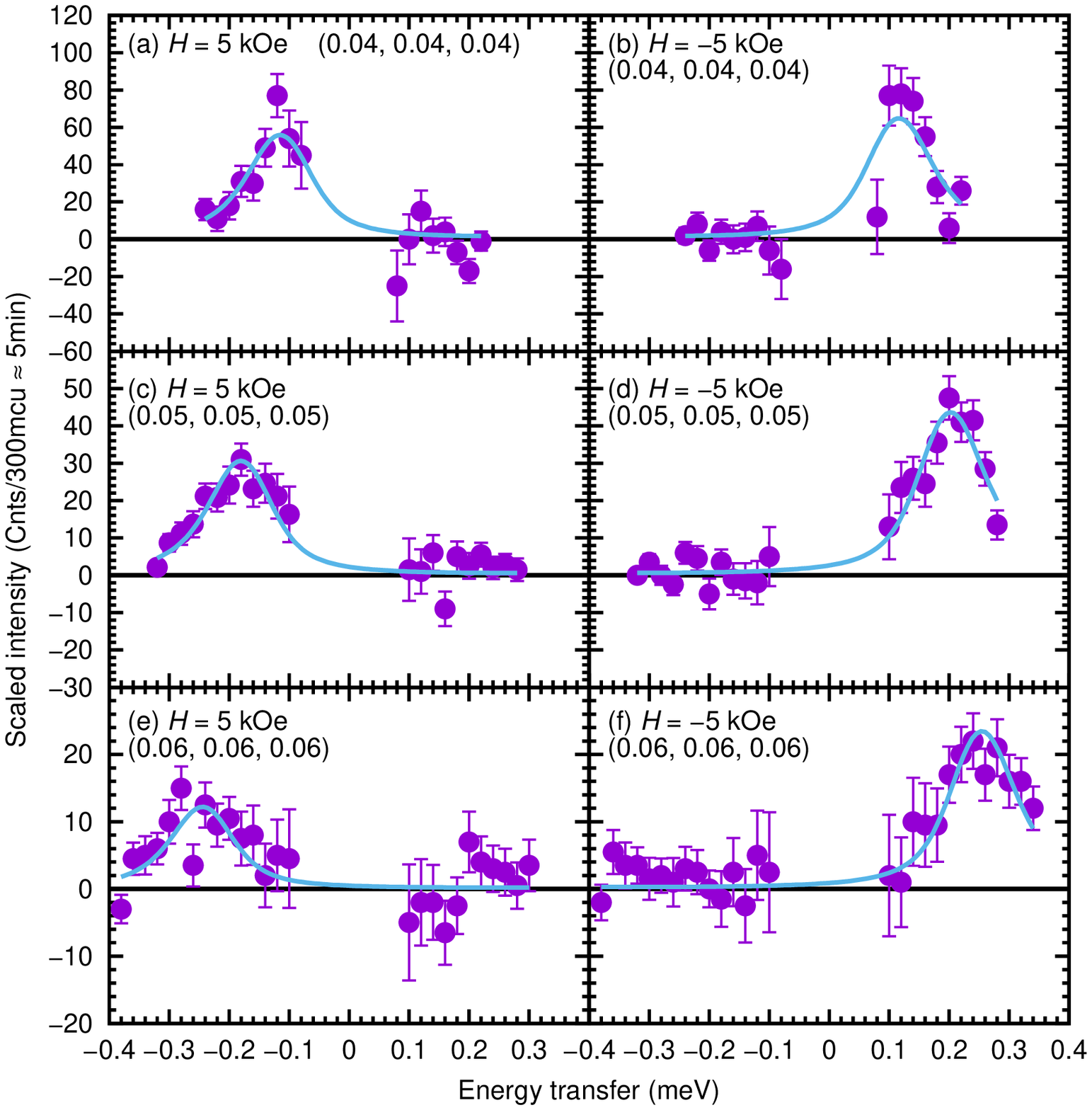}
\includegraphics[scale=0.48, angle=-90, trim={10cm 3cm 0cm 0cm}]{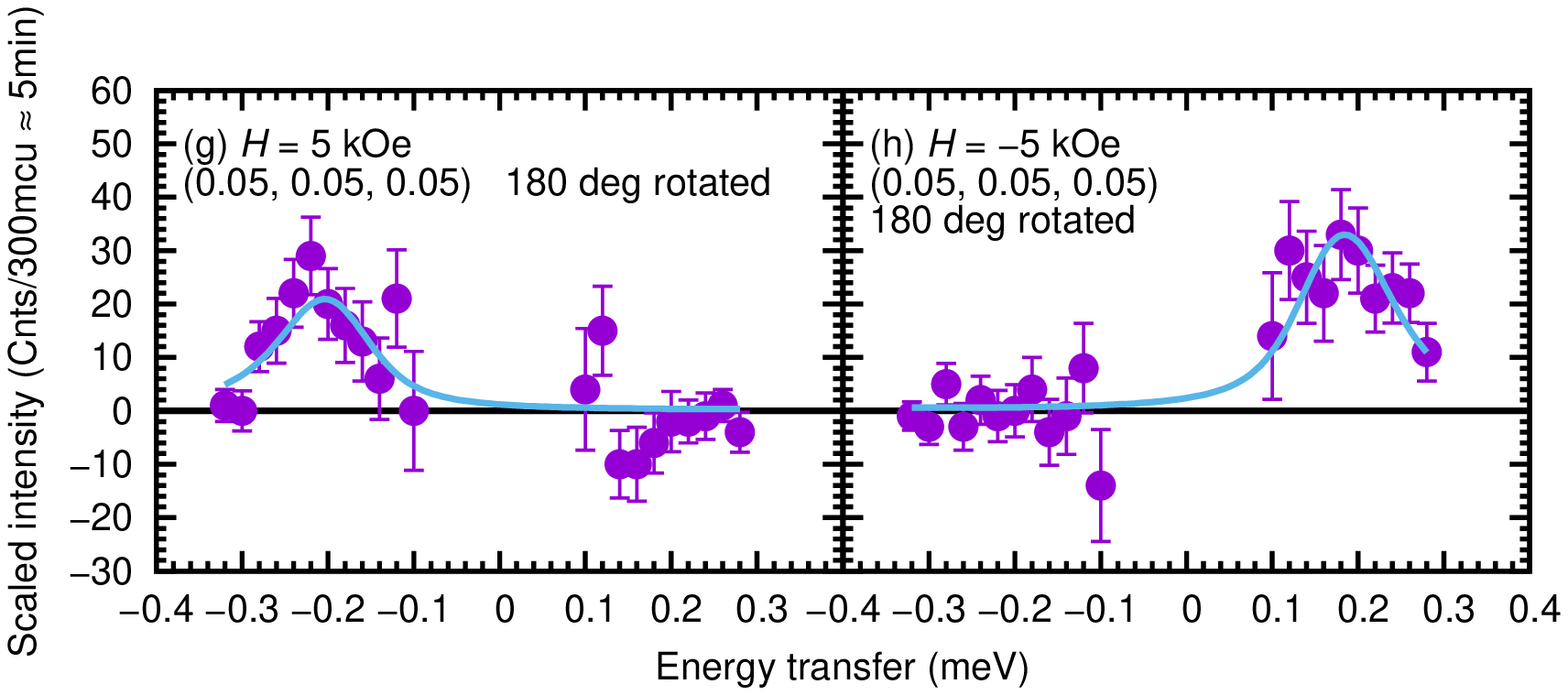}
\caption{(Color online) Representative constant-$\vec{Q}$ scans measured at $T = 27$~K.
(a) and (b) $h = 0.04$~(r.~l.~u.). 
(c) and (d) $h = 0.05$~(r.~l.~u.).
(e) and (f) $h = 0.06$~(r.~l.~u.).
The external field was set to $H = 5$~kOe for (a), (c), and (e), whereas the field was switched to $H = -5$~kOe for (b), (d), and (f).
(g) and (h) Constant-$\vec{Q}$ scans measured at $T = 27$~K and $h = 0.05$~(r.~l.~u.) with the sample rotated for 180 degrees.
The direction of magnetic field [5~kOe for (g) and -5~kOe for (h)] with respect to the incident neutron direction was unchanged.
For all the spectra, background was removed using the base temperature datasets; the data around the elastic position were discarded since huge elastic background results in unacceptable uncertainties.
The solid lines are the best fits to the model described in the text.
}
\end{figure}

Generally, the detailed balance law for the scattering function $S(\vec{Q}, E)$ is given as:
\begin{equation}
S(\vec{Q}, E) = {\rm e}^{E/k_{\rm B}T} S(-\vec{Q}, -E),
\end{equation}
where $k_{\rm B}$ is the Boltzmann constant, and ${\rm e}^{E/k_{\rm B}T} \simeq 1$ in our experimental condition.
Trivially, this becomes symmetric $S(\vec{Q}, E) \simeq S(\vec{Q}, -E)$ for centrosymmetric systems.
Therefore, the asymmetric appearance of the magnon peak can be naturally understood as a consequence of the noncentrosymmetricity of the underlying crystal.
As detailed below, this asymmetry of the inelastic spectra is a direct evidence of the asymmetry of the magnon dispersion relation, {\it i.e.} $E(q) \neq E(-q)$.

Shown in Figs.~3(g) and 3(h) are the inelastic spectra observed with the sample rotated for $\Delta \omega = 180^{\circ}$ with keeping magnetic-field and $\vec{Q}$ directions unchanged.
Note that this 180$^{\circ}$-rotation operation is identical to the simultaneous reversal of $H$ and $\vec{Q}$ with fixed sample $\omega$.
As is clearly seen in the figure, the magnon peak appears in the same side as those in the corresponding spectra shown in Figs.~3(c) and 3(d).
This confirms that the asymmetric appearance is related to the intrinsic chirality of the crystal structure, and is consistent with the theoretical expectation~\cite{MelcherRL73}.

To quantitatively estimate the excitation energy, each observed spectrum was fitted to the following scattering function convoluted with the instrumental resolution~\cite{ChesserNJ73}:
\begin{equation}\label{eq:lorentz}
  S(\vec{Q}, E)^{\pm} = \frac{I_{\rm SQ} \Gamma}{\Gamma^2 + [E \pm E(\vec{q})]^2},
\end{equation}
where $\pm$ stands for the sign of $H = \pm 5$~kOe.
For the resolution-convolution fitting, we need the dispersion relation in the resolution ellipsoid, and hence the asymmetric dispersion given as Eq.~(\ref{eq:dispersion}) is assumed.
We first performed the fitting to each inelastic spectrum with fixed $E_0 \simeq 0.02$~meV and $\vec{q}_0 \simeq (0.016, 0.016, 0.016)$, which are rough estimations obtained from raw peak positions.
The fitting results are shown in the corresponding figures by the solid lines; the observed spectra are well reproduced by the resolution convoluted Eq.~(\ref{eq:lorentz}).
The magnon energy at each $\vec{Q}$-position was estimated from the obtained $D_{\rm s}$ parameter, and is plotted as a function of $\vec{q} (= \vec{Q})$ in Fig.~4.
The clear shift of the magnon dispersion relation is seen in the figure.

In the continuous limit, a model Hamiltonian for MnSi may be written as~\cite{KataokaM87,ChizhikovVA12}:
\begin{equation}\label{eq:Hamiltonian}
{\cal H} = \int {\rm d}\vec{r} 
\left [ \frac{J}{2a} (\nabla \vec{S})^2 - \frac{D}{a^2}\vec{S}\cdot(\nabla\times\vec{S})
  - \frac{K_1}{2a^3}\sum_i S_i^4 - \frac{h}{a^3}S_{\rm z} \right ],
\end{equation}
where $a$ and $\vec{S}$ are the lattice constant and continuous spin field, respectively.
The first, second, third and the last terms stand for the exchange interaction (exchange parameter: $J$), DM interaction (DM parameter: $D$), cubic anisotropy (anisotropy parameter: $K_1$), and Zeeman energy (reduced external field: $h = -g \mu_{\rm B} H_{\rm ext}$, where $g$, $\mu_{\rm B}$, and $H_{\rm ext}$ are the $g$-factor, Bohr magneton and external field, respectively).
The suffix $i = x, y, z$ stands for the orthogonal coordinate axis, where the $z$-axis is defined along the reduced external field, {\it i.e.}, $\vec{z} \parallel \vec{h}$.
The dispersion relation for the magnon excitations along $\vec{q} \parallel \vec{h}$ is indeed written in the form of Eq.~(\ref{eq:dispersion}) with the  relations: $D_{\rm s} = JSa^2$, $q_0 = -D/aJ$, and $E_0 = h - h_{\rm c} = h - ( D^2S/J - 4 K_1S^3/3)$, where $S$ is the spin size and $h_{\rm c}$ is the critical field, above which the induced-ferromagnetic phase is stabilized.
In the above theory, the bottom of the quadratic dispersion, $q_0$, is given as $-D/aJ$, and accordingly the shift direction by the sign of the DM interaction $D$.
The earlier polarized-neutron-diffraction study concluded $D > 0$ from the anticlockwise nature of the zero field helical structure~\cite{IshidaM85}.
For this sign of $D$, the direction of the dispersion shift is expected to be antiparallel to $\vec{h}$, {\it i.e.}, $q_0 < 0$ for $h > 0$.
It should be further noted that $q_0$ corresponds to the modulation vector of the zero-field helical structure in the mean-field level~\cite{Nakanishi80}.

\begin{figure}
\includegraphics[scale=0.4, angle=-90]{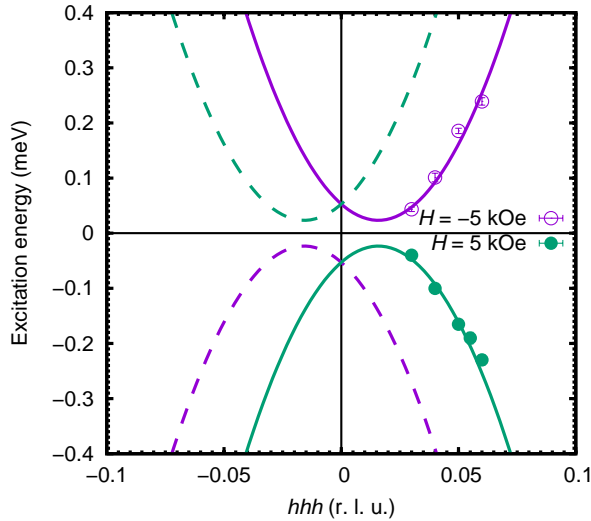}
\caption{(Color online) Obtained magnon dispersion relations in the induced-ferromagnetic phase of MnSi.
Open circles represent the excitation energies for $H = -5$~kOe, whereas filled circles for $H = 5$~kOe.
Magnon excitation energies were estimated from the inelastic spectra using resolution convoluted fitting to the Lorentzian-type function described in the text.
Solid lines stand for the shifted quadratic dispersions for $H = \pm 5$kOe given by Eq.~(\ref{eq:dispersion}) with the parameters $D_{\rm s} = 21(1)$~meV\AA$^2$, $E_0 = 0.02(1)$~meV.
Dotted lines are expected dispersion relations centered at $Q = - |q_0|$.
}
\end{figure}

The experimentally obtained magnon dispersion relation is fitted to the shifted quadratic dispersion relation Eq.~(\ref{eq:dispersion}) with $D_{\rm s}$ and $E_0$ as adjustable parameters.
We assume $q_0 = 0.038$~\AA$^{-1}$ $= |(0.016, 0.016, 0.016)|$~(r.l.u.) as estimated from the modulation period of the zero-field helical structure, since it cannot be reliably obtained in the dispersion-relation fitting because of the limited $\vec{Q}$-range for this small angle neutron inelastic scattering.
The optimum fitting result is shown in Fig.~4, where satisfactorily agreement is seen between the observed and model dispersion relations.
This also confirms the validity of the $q_0$ assumption.
The gap energy was estimated as $E_0 = 0.02(1)$~meV, which is comparable to the expected gap $h - h_{\rm c} \sim 0.014$~meV, as we measured the magnon dispersion at $H = \pm 5$~kOe which is approximately 1.2~kOe away from the critical field.
The estimated $D_{\rm s} = 39(3)$~meV/(r.l.u.$^2$) = $21(1)$~meV\AA$^{2}$, which is in good agreement with the earlier estimation~\cite{BoniP11}.
Using $D_{\rm s}$ and $q_0$, $DS$ at $T = 27$~K may be estimated as $D S = 0.17(1)$~meV. 
It may be noted that this value is considerably smaller than the value reported earlier~\cite{Grigoriev09}, since in this study $DS$ is estimated at the high temperature, where the effective spin size may be strongly reduced.

The direction of the magnon shift is consistent with the above theoretical expectation for $D > 0$, as our experimental $\vec{H}$ is defined antiparallel to $\vec{Q}$, and consequently parallel to $\vec{h}$ in Eq.~(\ref{eq:Hamiltonian}).
The dotted lines in the figure are dispersion relations centered at $Q =  -|q_0|$ expected from the detailed balance law.
Although the negative $Q$ and larger $|E|$ ranges were not accessible in the present experimental setup, this clearly shows that the shift direction can be reversed by switching the external field direction.

The observed asymmetry in the excitation spectrum can be explained using the shifted magnon dispersion.
Exemplified by the $H = -5$~kOe case, the magnon creation energy (shown by the magenta solid line) is accessible in the observed $Q$ range for the positive $E$ side.
On the other hand, in the negative $E$ side the magnon annihilation provides much larger energy (shown by the magenta dotted line), which is out of the energy range of the present inelastic experiment.
The asymmetric appearance of the magnon peak is the direct evidence of the magnon dispersion shift.

The present confirmation of the shifted magnon dispersion brings about intriguing implication for manipulating the spin-wave-spin current in bulk ferromagnets.
Although the net spin current cancels out for the thermally populated magnons, by selectively exciting magnons with wavenumbers $\pm q$ we will have different group velocities $\vec{v}_{\vec{q}} = \partial \omega_{\vec{q}} / \partial \vec{q}$.
The nonreciprocal propagation of electrostatic spin waves~\cite{IguchiY15,SekiS15} is one of the phenomena originating from the different group velocities, and now we can expect the same nonreciprocal propagation for the microscopic ferromagnetic magnons.
The present confirmation further ensures that another interesting effect, the magnon-mediated DM torque originally proposed in the ferromagnetic thin films~\cite{Manchon14}, may be realized in the bulk ferromagnet.
Further study in this direction may be quite interesting.

\section{Summary}
In summary, we have performed the small angle inelastic neutron scattering experiment in the noncentrosymmetric binary compound MnSi.
We have directly confirmed the shifted magnon dispersion in the bulk noncentrosymmetric ferromagnet in a microscopic length scale for the first time.
It was further shown that the magnon dispersion shift can be reversed by switching the external magnetic field direction.

\begin{acknowledgments}
The authors thank N. Nagaosa, Y. Endoh, and K. Kakurai for stimulating discussions.
This work was partly supported by Grant-In-Aid for Scientific Research (24224009) from MEXT of Japan.
The work at HFIR, Oak Ridge National Laboratory, was sponsored by the Division of Scientific User Facilities, Office of Basic Energy Science, US Department of Energy (DOE), and was partly supported by the US-Japan Collaborative Program on Neutron Scattering.
\end{acknowledgments}

\end{document}